\documentclass[preprintnumbers,prd,twocolumn,amsmath,amssymb,nofootinbib,superscriptaddress]{revtex4-2}

\usepackage{graphicx}
\usepackage{color}
\usepackage{bm}

\newcommand{\beq}{\begin{equation}}
\newcommand{\eeq}{\end{equation}}
\newcommand{\bea}{\begin{eqnarray}}
\newcommand{\eea}{\end{eqnarray}}
\newcommand{\bseq}{\begin{subequations}}
\newcommand{\eseq}{\end{subequations}}

\newcommand{\rf}[1]{(\ref{#1})}

\begin{document}

\title{A `singular' bounce in the theory of gravity with non-minimal derivative coupling}

\author{Sergey V. Sushkov} 
\email{sergey_sushkov@mail.ru}
\affiliation{Institute of Physics, Kazan Federal University, Kremliovskaya street 16a, Kazan 420008, Russia}

\author{Rafkat G. Galeev}
\email{rafgaleev3@gmail.com}
\affiliation{Institute of Physics, Kazan Federal University, Kremliovskaya street 16a, Kazan 420008, Russia}

\begin{abstract}
We explore bounce scenarios in the framework of homogeneous and isotropic cosmological models with arbitrary spatial curvature in the theory of gravity with non-minimal derivative coupling of a scalar field to the curvature given by the term $(\zeta/H_0^2) G^{\mu\nu}\nabla_\mu\phi \nabla_\nu\phi$ in the Lagrangian. 
In general, a cosmological model is determined by six dimensionless parameters: the coupling parameter $\zeta$, and density parameters $\Omega_0$ (cosmological constant), $\Omega_2$ (spatial curvature term), $\Omega_3$ (non-relativistic matter), $\Omega_4$ (radiation), $\Omega_6$ (scalar field term).
%
As expected, we find that there are no turning points and/or bounces in cosmological models with negative or zero spatial curvature. At the same time, both a turning point and a bounce can exist in the model with positive spatial curvature. 
In particular, the bounce -- when the Universe's contraction is replaced by expansion  -- is happened at $\tau=\tau_*$ when $a(\tau_*)=a_{min} =(3\zeta\Omega_2)^{1/2}$, where $\tau=H_0 t$ is a dimensionless cosmic time and $\tau_*$ is a moment when the bounce happens. 
It is important fact that the value $a_{min}$ depends {\em only} on $\zeta$ and $\Omega_2$, and does {\em not} depend on $\Omega_0$, $\Omega_3$ and $\Omega_4$. 
%
We find that near the bounce $a(\tau)\approx a_{min}(1+\Delta\tau^2/18\zeta)$ and $h(\tau)\approx \Delta\tau/9\zeta$, where $\Delta\tau=\tau-\tau_*$. 
Thus, the scale factor $a(\tau)$, the Hubble parameter $h(\tau)$, and all corresponding geometrical invariants have a regular behavior near the bounce. In particular, $a(\tau)\to a_{min}$, $h(\tau)\to 0$ as $\tau\to \tau_*$. As well the values characterizing matter energy densities, such as $\rho_m\sim a^{-3}$ and $\rho_r\sim a^{-4}$, are regular near the bounce.  
Nevertheless, though the spacetime geometry and energy densities remain to be regular near the bounce, the scalar field has a singular behavior there. Namely, 
$\phi'\propto 1/\Delta\tau^2 \to\infty$ as $\Delta\tau\to 0$.
As a result, we conclude that the complete dynamical system describing the cosmological evolution in theory of gravity with non-minimal derivative coupling
is singular near the bounce. On our knowledge, such the scenario, when the spacetime geometry and matter energy densities remain to be regular at approaching the universe evolution to the moment of bounce, while the behavior of scalar field becomes singular, was unknown before. For this reason, we term this scenario as a {\em `singular' bounce}.

\end{abstract}

 \maketitle

\section{Introduction}
One of the long-standing fundamental challenges of theoretical cosmology is the problem of an initial singularity. This problem first appeared within the Friedmann cosmology for the simplest cosmological models \cite{Friedman_1922,Friedman_1924}. Many years later, Penrose and Hawking had formulated the cornerstone singularity theorems of General Relativity \cite{Penrose_1965,Hawking_1967,HP_1970} (for some historical details see e.g. \cite{SenovillaGarfinkle, Steinbauer}).
The Hawking-Penrose theorems comprise a body of rigorous results in Lorentzian differential geometry that, under physically reasonable conditions, imply the occurrence of a “singularity” in the sense of causal geodesic incompleteness of the spacetime manifold. 
In particular, Hawking's singularity theorem is for the whole universe, and works backwards in time: it guarantees that the (classical) Big Bang has infinite density and a 3d volume of the universe becomes zero. 
The basic supposition of the theorem when it only holds is that matter filling the universe obeys so called the strong energy condition. This is quite reasonable, since one can expect that all kinds of matter in the nature must fulfill the strong energy condition.

A ``bounce'' is not a singular but a regular scenario of cosmological evolution. During the bounce a contraction of the universe (if go backwards in time) is changed by an expansion. At the same time, the volume of the universe remains to be non zero, and all physical and geometrical characteristics remains to be finite. To provide such the scenario and overcome the predictions of the Hawking-Penrose theorems, one should either to suppose existing and dominating at early stage of the Universe's evolution of some exotic (in fact anti-gravitating) matter violating standard energy conditions, or to modify somehow general relativity.  


To date, many different versions of modified or extended theories of gravity have been proposed (see surveys 
\cite{Review_Salvatore:2011, Review_Clifton_etal, Review_ModGrav:2013, Review_Berti_etal, Review_Nojiri:2017, Review_Langlois:2019} and references therein).
And many modified theories of gravity describe bouncing cosmological models (see a modern, complete and detailed review of various bouncing scenarios in modified theories of gravity given in the Ph.D. thesis \cite{Agrawal:2024gbm}).
One of models intensively studied today is Horndeski theory of gravity \cite{Horndeski} derived in the 1970s as an attempt to obtain the most general action for a scalar-tensor theory with a single scalar degree of freedom and second-order field equations. In 2011 Horndeski gravity has been rediscovered in the context of generalized Galileon theories \cite{Kobayashi:2011}, and since the interest in this model has only growing.\footnote{The literature dedicated to various aspects of Horndeski gravity is very vast, and its survey lays out of the scope of this work. The reader interesting in this topic can find some references in the already mentioned surveys \cite{Review_Berti_etal, Review_Clifton_etal}.}

The important subclass of Horndeski gravity is represented by models with a non-minimal derivative coupling of a scalar field $\phi$ with the Einstein tensor with the action
\bea
	S &=& \displaystyle\frac12\int d^4x\sqrt{-g}\,
	\bigg[\frac{1}{8\pi} (R-2\Lambda) 
	\nonumber\\
	&& -\left( g^{\mu\nu}+\eta G^{\mu\nu} \right)\nabla_{\mu}\phi\nabla_{\nu}\phi\bigg]+S^{(m)},
	\label{action}
\eea
where $R$ is the scalar curvature, $G_{\mu\nu}$ is the Einstein tensor, $\Lambda$ is the cosmological constant, 
and $S^{(m)}$ is the action for ordinary matter fields, supposed to be minimally coupled to gravity in the usual way.
Additionally, the theory (\ref{action}) 
involves the coupling parameter $\eta$ with dimension of ({\em length})$^2$, which leads to interesting features of astrophysical objects. In particular, black holes \cite{Rinaldi:2012, Minamitsuji:2013, Anabalon:2014, Babichev:2014, Kobayashi:2014, Babichev:2015}, wormholes \cite{Sushkov:2012b, Sushkov:2014}, and neutron stars \cite{Rinaldi:2015, Rinaldi:2016, Silva:2016, Maselli:2016, Eickhoff:2018, KasSus:2023} have been widely explored within this theory.   		
As well, the non-minimal derivative coupling leads to very interesting cosmological consequences, which have been intensively studied in our recent works
\cite{Sushkov:2009, SarSus:2010, Sushkov:2012, SkuSusTop:2013, MatSus:2015, StaSusVol:2016, StaSusVol:2019, Galeev_etal:2021, Sushkov:2023aya}.

In Ref. \cite{Sushkov:2023aya} we have explored homogeneous and isotropic cosmological models with arbitrary spatial curvature in the theory (\ref{action}) and found that in the case of the positive spatial curvature the Hubble parameter turns to zero at some small minimal value of $a=a_{min}\approx (3\zeta\Omega_2)^{1/2}$, where $\zeta=\eta H_0^2$ is a dimensionless coupling parameter, $\Omega_2=k/{\rm a}_0^2 H_0^2$ is a dimensionless curvature density parameters ($k=+1$), and ${\rm a}_0$, $H_0$ are values of the scale factor ${\rm a}(t)$ and the Hubble parameter $H(t)$ at the present moment of time $t_0$. Such the behavior of $H(t)$ indicates the possibility of a bounce scenario in the model. However, the study of the bounce scenario in Ref. \cite{Sushkov:2023aya} was not a main task of the work, and results concerning the bounce was only preliminary   

The aim of this work is to explore in details the bounce scenario in the theory (\ref{action}) in homogeneous and isotropic cosmological models with arbitrary spatial curvature.



\section{Field equations\label{II}}
%
Varying the action \rf{action} with respect to $g_{\mu\nu}$ and $\phi$ gives the field equations, respectively:
\bseq\label{fieldeq}
\bea
\label{eineq}
&& G_{\mu\nu}=-g_{\mu\nu}\Lambda+8\pi\big[T_{\mu\nu}^{(m)}+T_{\mu\nu}^{(\phi)}
+\eta \Theta_{\mu\nu}\big], \\
\label{eqmo}
&&[g^{\mu\nu}+\eta G^{\mu\nu}]\nabla_{\mu}\nabla_\nu\phi=0,
\eea
\eseq
where $T^{(m)}_{\mu\nu}$ is a stress-energy
tensor of ordinary matter, and
\bea \label{T}
T^{(\phi)}_{\mu\nu}&=&\nabla_\mu\phi\nabla_\nu\phi-
{\textstyle\frac12}g_{\mu\nu}(\nabla\phi)^2, \\
\Theta_{\mu\nu}&=&-{\textstyle\frac12}\nabla_\mu\phi\,\nabla_\nu\phi\,R
+2\nabla_\alpha\phi\,\nabla_{(\mu}\phi R^\alpha_{\nu)}
\nonumber\\
&&
+\nabla^\alpha\phi\,\nabla^\beta\phi\,R_{\mu\alpha\nu\beta}
+\nabla_\mu\nabla^\alpha\phi\,\nabla_\nu\nabla_\alpha\phi
\nonumber\\
&&
-\nabla_\mu\nabla_\nu\phi\,\square\phi-{\textstyle\frac12}(\nabla\phi)^2
G_{\mu\nu}
\label{Theta}\\
&&
+g_{\mu\nu}\big[-{\textstyle\frac12}\nabla^\alpha\nabla^\beta\phi\,
\nabla_\alpha\nabla_\beta\phi
+{\textstyle\frac12}(\square\phi)^2
\nonumber\\
&& \ \ \ \ \ \ \ \ \ \ \ \ \ \ \ \ \ \  \ \ \   \ \ \  \ \ \ \ \ \
\ \ \ \ \ -\nabla_\alpha\phi\,\nabla_\beta\phi\,R^{\alpha\beta}
\big]. \nonumber
\eea
Due to Bianchi identity $\nabla^\mu G_{\mu\nu}=0$ and the conservation law
$\nabla^\mu T^{(m)}_{\mu\nu}=0$, Eq. \rf{eineq} leads to the differential
consequence
\beq
\label{BianchiT}
\nabla^\mu\big[T^{(\phi)}_{\mu\nu}+\eta \Theta_{\mu\nu}\big]=0.
\eeq
Substituting Eqs. \rf{T} and \rf{Theta} into \rf{BianchiT}, one can check straightforwardly that the differential consequence \rf{BianchiT} is equivalent to \rf{eqmo}.  In other words, Eq. \rf{eqmo} is a differential consequence of Eq. \rf{eineq}.

Let us consider the Friedmann – Lema\^{i}tre – Robertson – Walker cosmological models with the metric
\begin{equation}
\label{metric} 
ds^2=-dt^2+\textrm{a}^2(t)\left[\frac{dr^2}{1-kr^2}+r^2(d\theta^2 +\sin^2\theta d\varphi^2)\right],
\end{equation}
where $k=0,\pm1$, $\textrm{a}(t)$ is the scale factor, and $H(t)=\dot{\textrm{a}}(t)/\textrm{a}(t)$ is the Hubble parameter. 
Denoting the present moment of time as $t_0$, we have $\textrm{a}_0=\textrm{a}(t_0)$ and $H_0=H(t_0)$. 
Supposing homogeneity and isotropy, we also get $\phi=\phi(t)$ and $T^{(m)}_{\mu\nu}=\mathop{\rm diag}(\rho,p,p,p)$, where $\rho=\rho(t)$ is the energy density and $p=p(t)$ is the pressure of matter.


The general field equations \rf{fieldeq} written for the metric \rf{metric}
give the following two independent equations:
\bseq\label{genfieldeq}
\bea
  \label{00cmpt}
  &&3\left(H^2+\frac{k}{\textrm{a}^2}\right) =\Lambda +8\pi\rho
  \nonumber\\
  &&\qquad\qquad\qquad\quad
    +4\pi{\psi}^2\left(1-9\eta \left(H^2+\frac{k}{3\textrm{a}^2}\right)\right),
  \\
  \label{eqmocosm}
  &&\dot\phi\left(1-3\eta \left(H^2+\frac{k}{\textrm{a}^2}\right)\right)=\frac{Q}{\textrm{a}^3}, 
\eea
\eseq
where the dot means a derivative with respect to $t$.
Here Eq. \rf{00cmpt} is the modified Friedmann equation, i.e. the $tt$-component of \rf{eineq}, while Eq. \rf{eqmocosm} is the first integral of the scalar field equation \rf{eqmo}, where $Q$ is a constant of integration associated with the scalar charge.

Assume that the matter filling the universe is a mixture of a radiation and a non-relativistic component:
\beq\label{matter}
\rho=\rho_m+\rho_r=\rho_{{m}0} \left(\frac{\textrm{a}_0}{\textrm{a}}\right)^{3}
+\rho_{{r}0} \left(\frac{\textrm{a}_0}{\textrm{a}}\right)^{4}.
\eeq


Now let us introduce the dimensionless time $\tau$, scale factor $a$, Hubble parameter $h$, coupling parameter $\zeta$ and scalar charge $q$ as follows: 
\beq\label{param1}
\tau=H_0 t, \
a=\frac{\textrm{a}}{\textrm{a}_0}, \ 
h=\frac{H}{H_0}, \
\zeta=\eta H_0^2, \
q=\frac{Q}{\textrm{a}_0^3 H_0},
\eeq
and the following dimensionless density parameters:
$$
\Omega_{0}=\frac{\Lambda}{3H_0^2}, \quad
\Omega_{2}=\frac{k}{\textrm{a}_0^2 H_0^2}, \quad
\Omega_{3}=\frac{\rho_{m0}}{\rho_{cr}}, \quad
$$
\beq\label{param2}
\Omega_{4}=\frac{\rho_{r0}}{\rho_{cr}}, \quad
\Omega_{6}=\frac{4\pi Q^2}{3\textrm{a}_0^6 H_0^2},
\eeq
where $\rho_{cr}=3H_0^2/8\pi$ is the critical density. We will assume in this work that $\Lambda \ge 0$, hence $\Omega_0$ is always not negative, i.e. $\Omega_0\ge 0$. The parameter $\Omega_2$ is defined such that its sign is the same as that of $k$.\footnote{Note that the dimensionless curvature density parameter is usually defined as $\Omega_{k}=-\frac{k}{\textrm{a}_0^2 H_0^2}$, which has the opposite sign compared to $\Omega_2$.} 
Here it is also worth to emphasize the physical meaning of the dimensionless coupling parameter $\zeta$. 
The parameter $\eta$ has the dimension $(length)^2$, and so it will be convenient to use the notation $\eta=\varepsilon \ell^2$, where $\varepsilon$ is the sign of $\eta$, i.e. $\varepsilon=\pm 1$, and $\ell$ is a characteristic length which characterizes the nonminimal derivative coupling between the scalar field and curvature. The value $H_0$ determines the size of Hubble horizon as 
$ 
\ell_H={1}/{H_0}.
$ 
Hence, $\zeta$ is proportional to the square of ratio of two characteristic scales:
\beq
\zeta=\varepsilon \left(\frac{\ell}{\ell_H}\right)^2.
\eeq

In the dimensionless values, the equation (\ref{eqmocosm}) reads
\beq\label{eqmodimless}
\phi'\left(1-3\zeta\left(h^2+\frac{\Omega_2}{a^2}\right)\right)=\frac{q}{a^3},
\eeq
where a prime means a derivative with respect to $\tau$. 
Now, substituting $\dot\phi=H_0\phi'$ from Eq. \rf{eqmodimless} into \rf{00cmpt}, we can rewrite the modified Friedmann equation in terms of dimensionless values:
\beq\label{geneqH}
h^2=\Omega_{0}
-\frac{\Omega_2}{a^2} 
+\frac{\Omega_3}{a^3} 
+\frac{\Omega_4}{a^4}
+\frac{\Omega_{6}\big(1-3\zeta (3h^2+\frac{\Omega_2}{a^2})\big)}
{a^6 \big(1-3\zeta (h^2 +\frac{\Omega_2}{a^2})\big)^2}.
\eeq
For given model parameters $\zeta$ and $\Omega_i$, Eq. \rf{geneqH} completely determines the scale factor $a(\tau)$ and hence the whole cosmological evolution of the Universe. It is necessary noticing that the parameters are not independent. 
Really, at $\tau=\tau_0$ one has $a_0=1$ and $h_0=1$, then Eq. \rf{geneqH} reduces to
\beq\label{constr}
1=\Omega_{0}-\Omega_2+\Omega_{3}+\Omega_{4} 
+\frac{\Omega_{6}\big(1-3\zeta(3+\Omega_2)\big)}{\big(1-3\zeta(1+\Omega_2)\big)^2}.
\eeq
The latter represents a constraint relating values of parameters $\Omega_{0}$, $\Omega_2$, $\Omega_{3}$, $\Omega_{4}$, and $\Omega_{6}$ at the present time. 
For practical purposes, it will be convenient to rewrite the constraint \rf{constr} as follows
\beq\label{omega6}
\Omega_6=\frac{\big(1-3\zeta(1+\Omega_2)\big)^2}{1-3\zeta(3+\Omega_2)}\,
(1-\Omega_0+\Omega_2-\Omega_3-\Omega_4).
\eeq 
Thus, one has five independent parameters $\zeta$, $\Omega_0$, $\Omega_2$, $\Omega_3$, $\Omega_4$ with additional requirements: $\zeta\ge 0$, $\Omega_0\ge 0$, $\Omega_3\ge 0$, $\Omega_4\ge 0$, and $\Omega_6\ge 0$.

Denoting $y=h^2$ and bringing all terms in \rf{geneqH} to the common denominator
yields
\beq\label{cube}
\frac{P(a,y)}{\big(1-3\zeta (y +\frac{\Omega_2}{a^2})\big)^2}=0,
\eeq
where
\beq
P(a,y)=9\zeta^2 y^3+c_2(a)y^2+c_1(a)y+c_0(a)
\eeq
is the cubic in $y$ polynomial with the coefficients
\bea
c_2(a) &=& -6\zeta\left(1-\frac{3\zeta\Omega_2}{a^2}\right) -9\zeta^2\left(\Omega_0-\frac{\Omega_2}{a^2}+\frac{\Omega_3}{a^3}+\frac{\Omega_4}{a^4}\right),
\nonumber\\
c_1(a) &=& 
\left(1-\frac{3\zeta\Omega_2}{a^2}\right)^2 
\nonumber\\
&&
+6\zeta \left(1-\frac{3\zeta\Omega_2}{a^2}\right) \left(\Omega_0-\frac{\Omega_2}{a^2}+\frac{\Omega_3}{a^3}+\frac{\Omega_4}{a^4}\right)
\nonumber\\
&&
+\frac{9\zeta\Omega_6}{a^6},
\nonumber\\
c_0(a) &=& -\left(1-\frac{3\zeta\Omega_2}{a^2}\right)^2
\left(\Omega_0-\frac{\Omega_2}{a^2}+\frac{\Omega_3}{a^3}+\frac{\Omega_4}{a^4}\right) 
\nonumber\\
&&
-
\frac{\Omega_6}{a^6}\left(1-\frac{3\zeta\Omega_2}{a^2}\right)
\label{c0}
\eea
Eq.\rf{cube} will be fulfilled if $P(a, y) = 0$ and $1-3\zeta (y +\frac{\Omega_2}{a^2})\not= 0$, hence the problem reduces to studying roots of the cubic polynomial. Finding a particular root $y_i$, one determines the algebraic dependence of the Hubble parameter $h$ on the scale factor $a$. The relation to the physical time is then determined by the quadrature
\beq\label{quadrature}
\int_{a=1}^a \frac{d\tilde a}{\tilde a h(\tilde a)}=H_0(t-t_0).
\eeq

\section{Turning points and bounces in the Universe evolution}
A turning point in the Universe evolution may occur at a moment $t=t_{*}$, when the scale factor ${\rm a}(t)$ reaches either its maximal or minimal value. In the first case, the turning point means that the expansion of the Universe stops and is replaced by contraction. Such a scenario is natural, for example, for the closed Friedmann cosmological model, when the Universe is too massive and gravity is able to stop the expansion. In the second case, when ${\rm a}(t)$ reaches its minimal value, the situation is reversed, that is at the turning point Universe's contraction is replaced by expansion. Such a scenario, if possible, is usually called a cosmological {\em bounce}. It is worth noting that the Hawking singularity theorem in fact forbids bounces in the standard Big Bang cosmology. To construct a cosmological model with a bounce, one needs either to suppose existing and dominating at early stage of the Universe's evolution of some exotic anti-gravitating matter violating standard energy conditions, or to modify somehow general relativity.    

Let us consider if turning points exist in the theory of gravity with non-minimal derivative coupling. Necessary condition for the existence of an extremum (maximum or minimum) of $a(t)$ at a moment $t=t_*$ is $da(t_*)/dt=0$, 
or, equivalently, $H(t_*)=h(\tau_*)=y(a_*)=0$, where $a_*=a(\tau_*)$. Substituting $y(a_*)=0$ into (\ref{cube}), we get $c_0(a_*)=0$. Thus, a `bounce' solution $y=0$ of Eq. (\ref{cube}) can exist {\em if and only if} there exists $a_*$ such that $c_0(a_*)=0$. Using Eq. (\ref{c0}), we obtain two separate algebraic conditions for $a_*$:
\beq\label{condamax}
\left(1-\frac{3\zeta\Omega_2}{a_*^2}\right) \left(\Omega_0-\frac{\Omega_2}{a_*^2}+\frac{\Omega_3}{a_*^3}+\frac{\Omega_4}{a_*^4}\right)
+\frac{\Omega_6}{a_*^6} = 0,
\eeq
and
\beq\label{condamin}
\left(1-\frac{3\zeta\Omega_2}{a_*^2}\right) = 0.
\eeq
First of all, it is necessary to notice that both algebraic equations (\ref{condamax}) and (\ref{condamin}) have {\em no} solutions in case $\Omega_2 \le 0$. Therefore, in cosmological models with negative or zero spatial curvature there are no turning points. On the other hand, in the case of positive spatial curvature, when $\Omega_2 > 0$, both equations (\ref{condamax}), (\ref{condamin}) have solutions, and therefore turning points can exist.    

Let us first analyze the condition (\ref{condamax}) assuming that $\Omega_2 > 0$. In the simplest case $\zeta=0$, $\Omega_0=\Omega_3=\Omega_4=0$, the relation (\ref{omega6}) reads $\Omega_6=1+\Omega_2$, and one easily obtains from Eq. (\ref{condamax}) 
\beq
a_*^2=
\left(\frac{\Omega_6}{\Omega_2}\right)^{1/2}=\left(1+\frac{1}{\Omega_2}\right)^{1/2}.
\eeq 
Supposing $\Omega_2\ll 1$, one has $a_*^2 \approx 1/\Omega_2^{1/2} \gg 
1$.
In general, the transcendental equation (\ref{condamax}) can be qualitatively analyzed graphically. 
In particular, in the case when $\zeta\not=0$, $\Omega_3\not=0$, $\Omega_4\not=0$, but $\Omega_0=0$, the condition (\ref{condamax}) {\em always} has a root $a_*$ such that $a_*^2 > 1/\Omega_2^{1/2} \gg 1$. In case $\Omega_0\not=0$, the condition (\ref{condamax}) could still have a root $a_*
\gg 1$ depending on particular values of all parameters $\zeta$ and $\Omega_i$.    
Thus, the condition (\ref{condamax}) provides the value $a_*\gg 1$ corresponding to the turning point of the Universe evolution which turns out at some distant future $\tau=\tau_*>\tau_0$, when the scale factor $a(\tau)$ achieves its maximal value $a_{max}=a_*$ and the Universe’s expansion is replaced by contraction. It is worth emphasizing once more that the value $a_{max}$ depends, in general, on all parameters $\zeta$ and $\Omega_i$, so that $a_{max}=a_{max}(\zeta,\Omega_0,\Omega_2,\Omega_3,\Omega_4)$.  

Now let us consider the second condition (\ref{condamin}), which easily gives the following relation:
\beq
a_*^2 =3\zeta\Omega_2.
\eeq 
Since $\zeta\ll 1$ and $\Omega_2\ll 1$, one can conclude that $a_{*}=(3\zeta\Omega_2)^{1/2}\ll 1$, and  therefore, the turning point determined by Eq. (\ref{condamin}) turns out at early stages of the Universe evolution, when the scale factor $a(\tau)$ achieves its minimal value $a_{min}\equiv a_*=(3\zeta\Omega_2)^{1/2}$. 
Moreover, it is very important to stress that the value $a_{min}$ depends {\em only} on $\zeta$ and $\Omega_2$, and does {\em not} depend on $\Omega_0$, $\Omega_3$ and $\Omega_4$. Following Ref. \cite{StaSusVol:2016}, one can say that the cosmological constant encoded in $\Omega_0$ and the material filling encoded in $\Omega_3$ and $\Omega_4$ are screened at early stages of the Universe evolution.

To describe more accurately the behavior of $a(\tau)$ and $h(\tau)$ in the vicinity of the turning point $\tau_*$, we will obtain a solution of Eq. (\ref{geneqH}) in the following form (for details see the appendix):
\bea
\label{asa}
a(\tau) &=& a_{min}\left(1+\frac{\Delta \tau^2}{18\zeta}\right) +O(\Delta \tau^4),
\\
\label{ash}
h(\tau) &=& \frac{\Delta \tau}{9\zeta} +O(\Delta \tau^3).
\eea
where with $\Delta \tau =\tau-\tau_*$.
Thus, $a(\tau)\to a_{min}$ and $h(\tau)\to 0$ as $\Delta\tau\to 0$, i.e. $\tau\to\tau_*$. 
The spacetime metric (\ref{metric}) near the turning point takes the following form
\beq
ds^2 \approx -dt^2 +{\rm a}_{min}^2\left(1+\frac{\Delta t^2}{9\eta}\right) ds_{(3)}^2,
\eeq
where ${\rm a}_{min}={\rm a}_0 a_{min}$, $\Delta t=t-t_*$, and $ds_{(3)}^2=dr^2/(1-r^2)+r^2(d\theta^2+\sin^2\theta d\varphi^2)$ is the metric of a unit 3d sphere. The corresponding invariant of a scalar curvature,
\beq
R=\frac{6({\rm a}_{min}^2 +9\eta)}{{\rm  a}_{min}^2(\Delta t^2+9\eta)},
\eeq 
is regular at the turning point $t=t_*$ and equal to 
$R_*=6\left(\frac{1}{{\rm  a}_{min}^2} +\frac{1}{9\eta}\right)$.\footnote{Note that in the limit $\eta\to\infty$, corresponding to the transition to general relativity, one obtains the standard expression $R_*=6/{\rm  a}_{min}^2$.} 
Ultimately, one can conclude that the moment $\tau_*$ in the Universe evolution is a bounce such that the scale factor reaches its minimal value $a_{min}=(3\zeta\Omega_2)^{1/2}$, the Hubble parameter turns out into zero, and the spacetime geometry remains to be regular.  

However, the above analysis is still incomplete, because it is necessary to consider as well a behavior of the scalar field near the bounce.
Substituting the asymptotics (\ref{asa}) and (\ref{ash}) into (\ref{eqmodimless}), we obtain
\beq\label{derphi}
\phi' \approx \frac{27\zeta q}{2a_{min}^3 \Delta \tau^2}.
\eeq
Thus, one has $\phi' \propto 1/\Delta\tau^2 \to \infty$ as $\Delta\tau\to 0$, i.e. $\tau\to\tau_*$. 
Therefore, despite the fact that the spacetime geometry is regular, the scalar field has a singular behavior near the bounce. 

\section{Concluding remarks}
In this paper we have explored bounce scenarios in the framework of homogeneous and isotropic cosmological models with arbitrary spatial curvature in the theory of gravity with non-minimal derivative coupling given by the action (\ref{action}). 
In general, the model depends on five independent dimensionless parameters: the coupling parameter $\zeta$, and density parameters $\Omega_0$, $\Omega_2$, $\Omega_3$, $\Omega_4$ (see Eqs. \rf{param1}, \rf{param2}), and a cosmological evolution is described by the modified Friedmann equation (\ref{geneqH}) together with the scalar field equation (\ref{eqmodimless}) and constraint (\ref{constr}).

First of all, as expected, we found that there are no turning points and/or bounces in cosmological models with negative or zero spatial curvature. At the same time, both a turning point and a bounce can exist in the model with positive spatial curvature. The turning point of the Universe evolution -- when the expansion is replaced by contraction -- is determined by the condition (\ref{condamax}). In the simplest case when $\zeta=0$, $\Omega_0=\Omega_3=\Omega_4=0$, the turning point corresponds to the maximal value of the scale factor, $a_{max}=\left(1+{1}/{\Omega_2}\right)^{1/4}$. Generally, the value $a_{max}$ depends on all parameters $\zeta$ and $\Omega_i$, so that $a_{max}=a_{max}(\zeta,\Omega_0,\Omega_2,\Omega_3,\Omega_4)$.  

The bounce -- when the contraction is replaced by expansion  -- is determined by the condition (\ref{condamin}), which easily gives $a_{min} =(3\zeta\Omega_2)^{1/2}$. 
Here it is necessary to stress the very important fact that the value $a_{min}$ depends {\em only} on $\zeta$ and $\Omega_2$, and does {\em not} depend on $\Omega_0$, $\Omega_3$ and $\Omega_4$. In Ref. \cite{StaSusVol:2016}, this feature was called as {\em screening} of the cosmological constant and matter at early stages of the Universe evolution. We found that near the bounce $a(\tau)\approx a_{min}(1+\Delta\tau^2/18\zeta)$ and $h(\tau)\approx \Delta\tau/9\zeta$, where $\Delta\tau=\tau-\tau_*$, and $\tau_*$ is a moment when the bounce happens. Thus, the scale factor $a(\tau)$, the Hubble parameter $h(\tau)$, and all corresponding geometrical invariants have a regular behavior near the bounce. In particular, $a(\tau)\to a_{min}$, $h(\tau)\to 0$ as $\tau\to \tau_*$. As well the values characterizing matter energy densities, such as $\rho_m\sim a^{-3}$ and $\rho_r\sim a^{-4}$, are regular near the bounce.  

Though the spacetime geometry and energy densities remain to be regular near the bounce, the scalar field has a singular behavior there. Really, we obtained that the time derivative of the scalar field is given by Eq.(\ref{derphi}) and diverges near the bounce as follows: $\phi'\propto 1/\Delta\tau^2 \to\infty$ as $\Delta\tau\to 0$.

As a result, we must conclude that the complete dynamical system represented by equations (\ref{eqmodimless}), (\ref{geneqH}), (\ref{constr}) is singular near the bounce. On our knowledge, such the scenario, when the spacetime geometry remains to be regular at approaching the universe evolution to the moment of bounce, while the behavior of scalar field becomes singular, was unknown before. For this reason, we can term this scenario as a {\em `singular' bounce}.

\section*{Acknowledgments}
{This work is supported by the Foundation for the Development of Theoretical Physics and Mathematics ``Basis'',  grant No. 24-1-1-39-1} and partially carried out in accordance with the Strategic Academic Leadership Program "Priority 2030" of the Kazan Federal University.

\section*{Appendix}
Let us obtain a `bounce' solution of the modified Friedmann equation (\ref{geneqH}). For this purpose, we recast Eq. (\ref{geneqH}) in the following form:
\beq\label{geneqy}
y=\Omega_{0}
-\frac{\Omega_2}{a^2} 
+\frac{\Omega_3}{a^3} 
+\frac{\Omega_4}{a^4}
+\frac{\Omega_{6}\big(1-9\zeta y -\frac{a_{*}^2}{a^2}\big)}
{a^6 \big(1-3\zeta y -\frac{a_{*}^2}{a^2}\big)^2},
\eeq
where we denote $y=h^2$ and $a_{*}^2=3\zeta\Omega_2$. Further, we will regard the above equation as an algebraic equation in respect to $y$ and will search its solution as the following series:
\beq\label{sery}
9\zeta y = y_1 \frac{\Delta a}{a_*} 
+y_2 \left(\frac{\Delta a}{a_*}\right)^2 
+y_3 \left(\frac{\Delta a}{a_*}\right)^3 + \dots,
\eeq
where $\Delta a=a-a_{*}$ and $y_n\ (n=1,2,3,\dots)$ are unknown coefficients. Substituting the series (\ref{sery}) into (\ref{geneqy}), one can find that
\beq
y_n=(-1)^{n+1}(n+1) +\tilde y_n, \quad n=1,2,3,\dots.
\eeq
where $\tilde y_1=0$ and the other $\tilde y_n\ (n=2,3,\dots)$ are proportional to $a_*^2$. For example,\footnote{Explicit expressions for $\tilde y_n\ (n=3,4,\dots)$ are quite cumbersome, and there is no reason to represent them here.} 
$$
\tilde y_2 = \frac{16 a_*^2}{9\Omega_6}\ (\Omega_4 +a_* \Omega_3 -a_*^2 \Omega_2 +a_*^4 \Omega_0).
$$
One can estimate $\tilde y_n$, using some reasonable values for parameters $\Omega_i$ and $\zeta$. Referring to Planck primary CMB data \cite{Planck:2018vyg}, we can consider the following values: $\Omega_0=0.6847$, $\Omega_2=0.044$, $\Omega_3=0.315$, $\Omega_4=0.0001$. As for $\zeta$, there are no evident observational data constraining its value. However, we can suppose that $\zeta$ is very small and take for estimation $\zeta=10^{-5}$. As the result, we find $\tilde y_2=2.45\times 10^{-8}$, $\tilde y_3=-5.31\times 10^{-9}$, etc, so that $\tilde y_n\ll 1\ (n=2,3,\dots)$. 
Neglecting $\tilde y_n$, we can recast Eq. (\ref{sery}) as follows
\beq\label{sery2}
9\zeta y \approx 2 \frac{\Delta a}{a_*} 
-3 \left(\frac{\Delta a}{a_*}\right)^2 
+4 \left(\frac{\Delta a}{a_*}\right)^3 + \dots,
\eeq
Keeping only the first order of $\Delta a$ and substituting $y=h^2=(\dot a/a)^2$, one has       
\beq
\left(\frac{\dot a}{a}\right)^2 = \frac{2}{9\zeta}\, \frac{a-a_*}{a_{*}}.
\eeq   
Integrating gives
\beq
a(\tau)=a_*\,\left(1+\tan^2((18\zeta)^{-1/2}\,\Delta \tau)\right),
\eeq
where $\Delta \tau=\tau-\tau_*$, and $\tau_*$ is a constant of integration. Taking into account the following terms in the expansion (\ref{sery2}), keeping the leading terms in the expansion in $\Delta\tau$ and substituting $a_*=a_{min}$, we eventually obtain the following asymptotics for $a(\tau)$ near the bounce
\beq
a(\tau)=a_{min}\,\left(1+\frac{\Delta\tau^2}{18\zeta}\right) +O(\Delta\tau^4).
\eeq   
For the Hubble parameter $h=\dot a/a$ one has
\beq
h(\tau) = \frac{\Delta \tau}{9\zeta} +O(\Delta \tau^3).
\eeq 




\end{document}